\documentclass[a4paper, amssymb, amsmath, reprint, aps, nofootinbib, twoside,pra]{revtex4-2}
\usepackage[utf8]{inputenc}
\usepackage{graphicx}
\usepackage{physics}
\usepackage{hyperref}
\begin{document}

% ================
% Begin Title Page
% ================

\title{Manipulating phonons of a trapped-ion system using optical tweezers}

\author{Yi Hong Teoh$^1$, Manas Sajjan$^1$, Zewen Sun$^1$, Fereshteh Rajabi$^1$, Rajibul Islam$^1$}
\affiliation{$^1$Institute for Quantum Computing and Department of Physics and Astronomy, University of Waterloo, 200 University Ave. West, Waterloo, Ontario N2L 3G1, Canada}

% ==============
% Begin Abstract
% ==============

\begin{abstract}
We propose an experimental architecture where an array of optical tweezers affords site-dependent control over the confining potential of a conventional radio-frequency ion trap.
The site-dependent control enables programmable manipulation of phonon modes of ions, with many potential applications in quantum information processing (QIP) and thermodynamics.
We describe protocols for programming the array of optical tweezers to attain a set of target phonon modes with high accuracy.
We propose applications of such controls in simulating quantum thermodynamics of a particle of programmable effective mass via Jarzynski's equality and improving the efficiency of sympathetic cooling and quantum logic gates in a multi-species ion system of disparate masses.
We discuss the required optical parameters in a realistic ion trap system and potential adverse effects of optical tweezers in QIP. 
Our scheme extends the utility of trapped-ions as a platform for quantum computation and simulation.
\end{abstract}

% ============
% End Abstract
% ============

\maketitle

% ==============
% End Title Page
% ==============

% ==================
% Begin Introduction
% ==================

\section{Introduction}
Collective vibrational or phonon modes of a laser-cooled trapped ion system, arising from the interplay between Coulomb repulsion and external trapping forces, provides unique advantages for quantum information processing (QIP).
Phonon-mediated interactions between internal qubit (generally qudit) states are at the heart of generating entangling quantum gates in quantum computation and quantum many-body Hamiltonians in quantum simulation \cite{Cirac1995quantum,Sorensen1999quantum,Zhu2006,kim2009entanglement,Korenblit2012quantum,Wright2017}. 
The ability to control these phonon modes would enhance the utility of the trapped ion system \cite{Olsacher2020scalable,shen2020scalable}.
In particular, quantum gates between ions of different masses, advantageous for scaling up quantum processors \cite{inlek2017multispecies,sosnova2020character}, can be made faster by manipulating their participation in the collective modes.
For maximum control over the phonon modes, site-dependent control over individual ions is necessary.
However, conventional radio-frequency ion traps offer only global or limited local control of the confining potential \cite{Mielenz2016arrays}.

Optical tweezers or tightly focused laser beams at each ion can potentially allow site-dependent manipulation of the confining potential through AC Stark effect.
In recent experiments, optical tweezers have indeed enabled bottom-up manipulation of a plethora of quantum systems \cite{Hou2013,Endres2016,Barredo2016,Barredo2018,omran2019generation,shen2020scalable}.
The ability to manipulate local potential of controllable strengths gives rise to the possibility of simulating mechanical properties of a system of disparate masses.
This also allows for investigation of controllable phonon-phonon interactions \cite{debnath2018observation}.
Further, fast optical switching, comparable to the conventional trap frequency, allows for probing phonon dynamics with possible applications in quantum thermodynamics. 
However, finding the desired optical tweezer strengths for a target set of phonon modes could be a challenging control problem.

In this work, we analytically and numerically investigate the extent to which trapped-ion phonon modes can be controlled by site-selective potentials afforded by optical tweezers, and a few potential applications of these capabilities.
We propose a non-linear optimization protocol to solve the optical tweezer control problem to attain a target set of normal mode frequencies within limits of realistic experimental errors.
On the other hand, we cast the problem of controlling the normal mode eigenvectors in the form of a set of linear problems.
We find that our control over the phonon modes is substantial when the optical tweezer trapping frequencies are comparable to the conventional trapping frequency.
As an application, we propose a concrete experimental setup that demonstrates the utility of Jarzynski's equality \cite{jarzynski1997nonequilibrium} for simulating quantum thermodynamics of a particle of programmable effective mass.
In another proposed application, we demonstrate that the control over eigenvectors can be leveraged to enhance coupling between ions of disparate masses, improving the efficiency of sympathetic cooling and mixed-species quantum logic gates \cite{inlek2017multispecies,sosnova2020character}.
In an example system consisting of $^{171}\mathrm{Yb}^+$ ions, we numerically verify that optical trapping frequencies of approximately $2\pi\times$1 MHz can be obtained with realistic experimental parameters.
However, the introduction of optical tweezers are accompanied by differential AC Stark effect and off-resonant scattering with potential implications in QIP experiments \cite{ozeri2007errors,islam2011onset}.
These effects can be minimized by choosing the frequency of the optical tweezer to be far-detuned from atomic transitions \cite{ozeri2007errors}.

% ================
% End Introduction
% ================

% ==========
% Begin Body
% ==========

\section{Effect of optical tweezers on normal modes}

In a conventional trapped ion system, ions are typically confined using a harmonic pseudopotential generated by RF and DC electrodes. 
The ions also experience mutual Coulomb repulsion.
In this paper, we consider an additional potential created by optical tweezers, i.e. tightly focused laser beams, on the ions.
Internal energy states of ions experience AC Stark shifts that follow the spatial profile of the optical intensity of tweezers.
Depending on the frequency of the tweezer light, the AC Stark shift will create either a trapping or an anti-trapping potential $\phi^{\mathrm{opt}}$.
Consequently, the total potential of the $N$-ion system $\phi^{\mathrm{total}}$ has the following form:
\begin{align}
\begin{split}
\phi^{\mathrm{total}} \left( \{ \mathbf{r}_i \}_{i=1}^N \right)
={}&\phi^{\mathrm{Coulomb}} \left( \{ \mathbf{r}_i \}_{i=1}^N \right) \\
& + \sum_i \phi^{\mathrm{conv}}(\mathbf{r}_i;M_i) \\
& + \sum_i \phi^{\mathrm{opt}}_{i}(\mathbf{r}_i), \label{eqn:totalpotential}
\end{split}
\end{align}
where,
\begin{equation}
\phi^{\mathrm{Coulomb}} \left( \{ \mathbf{r}_i \}_{i=1}^N \right)
=\sum_{i<j} \frac{q_i q_j}{4 \pi \varepsilon_0 \left\lVert \mathbf{r}_i
- \mathbf{r}_j\right\rVert}, \label{eqn:coulombpotential}
\end{equation}
and, 
\begin{align}
\begin{split}
\phi^{\mathrm{conv}}(\mathbf{r}_i;M_i,q_i)
= \frac{M_i}{2} \Bigl\{& [\omega_{x}^{\mathrm{conv}}(M_i,q_i)]^2 x_i^2 \\
& + [\omega_{y}^{\mathrm{conv}}(M_i,q_i)]^2 y_i^2 \\
& + [\omega_{z}^{\mathrm{conv}}(M_i,q_i)]^2 z_i^2 \Bigr\}. \label{eqn:harmonicpotential}
\end{split}
\end{align}
Here, $\mathbf{r}_i$, $M_i$ and $q_i$ are respectively the position vector, mass and electric charge of ion $i$, $\varepsilon_0$ is the permittivity of free space, and $\omega^{\mathrm{conv}}_{\alpha}(M_i,q_i)$ is the conventional mass and charge dependent trapping strength along the principal axis $\alpha \in \{x,y,z\}$ \cite{paul1990electromagnetic,wineland1997experimental,matteo2020quantum} (see Appendix~\ref{apdx:convtrap}).
We assume that $\omega^{\mathrm{conv}}_x \approx \omega^{\mathrm{conv}}_y \gg \omega^{\mathrm{conv}}_z$, such that the equilibrium configurations of the ions are in a linear chain along the $z$-direction \cite{Shmuel2008structural,Efrat2011quantum}.

\begin{figure*}[t]
    \centering
    \includegraphics[width=0.8\textwidth]{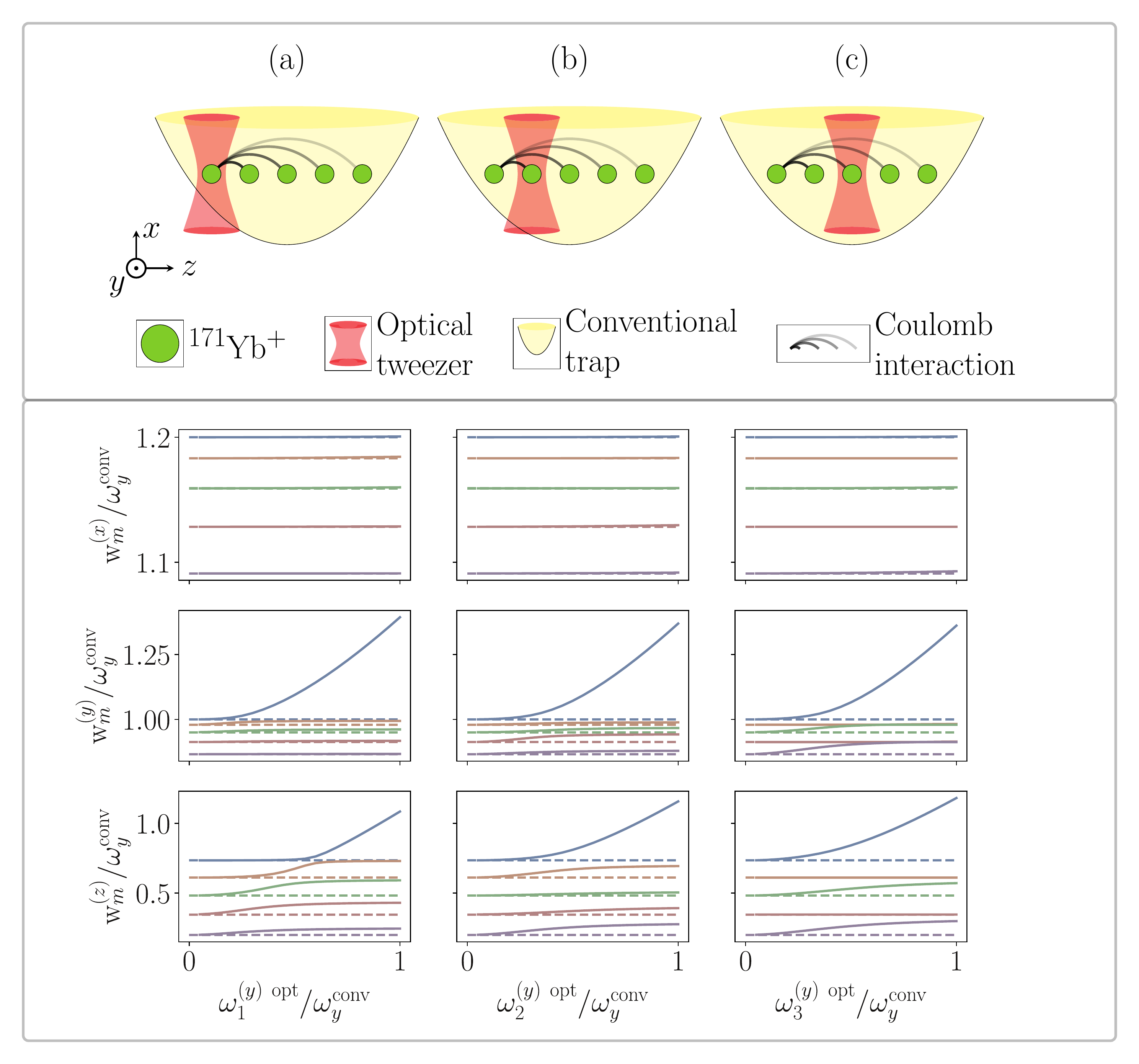}
    \caption{\textbf{Effects of an optical tweezer, with strengths $\omega^{(\alpha) \ \mathrm{opt}}_i$ ($\alpha=x,y,z$), focused onto ion-$i$ in a system with $N=5$ $^{171}\mathrm{Yb}^+$ ions (green filled circles).}
    Due to the reflection symmetry about the center ion, three independent tweezer locations are shown in (a) through (c).
    The ions are in a conventional harmonic trapping potential $\phi^{\mathrm{conv}}$ with trap strengths of, $\omega^{\mathrm{conv}}_x/2\pi = 1.2 \ \mathrm{MHz}$, $\omega^{\mathrm{conv}}_y/2\pi = 1 \ \mathrm{MHz}$ and $\omega^{\mathrm{conv}}_z/2\pi = 0.2 \ \mathrm{MHz}$.
    (Bottom panel) Calculated normal mode spectrum along $\alpha$, $w^{(\alpha)}_m$, changes with the trapping strength of the optical tweezer.}
    \label{fig:singletweezer}
\end{figure*}

For a Gaussian optical tweezer propagating along $x$, of intensity profile $I^{\mathrm{Gauss}}_i$, the optical potential $\phi^{\mathrm{opt}}_{i}$ on ion $i$ can be calculated from AC Stark shift formula \cite{Grimm200095}, 
\begin{align}
\begin{split}
\phi^{\mathrm{opt}}_{i}(\mathbf{r}_i)
={}&  -\sum_{a}\frac{3\pi c^2}{2\omega_{a,i}^3} \left( \frac{\Gamma_{a,i}}{\omega_{a,i} -\omega_{l,i}} + \frac{\Gamma_{a,i}}{\omega_{a,i} +\omega_{l,i}} \right) \\
& \times I^{\mathrm{Gauss}}_i(\mathbf{r}_i - \mathbf{r}^*_i), \label{eqn:generalOpticalPotential}
\end{split}
\end{align}
where $\Gamma_{a,i}$ and $\omega_{a,i} $ are the scattering rate and atomic transition frequency between the ground and excited state $a$ for ion $i$, respectively.
The summation in Eq.~(\ref{eqn:generalOpticalPotential}) is over all excited states, $\mathbf{r}_i^*$ is the equilibrium position of ion i, and
\begin{equation}
I^{\mathrm{Gauss}}_i(\mathbf{r}_i) = I_0^{(i)} \left( \frac{\sigma_0}{\sigma(x_i)} \right)^2 \mathrm{exp} \left( \frac{-2(y_i^2+z_i^2)}{\sigma^2(x_i)} \right) \label{eqn:gaussianbeam}. \\
\end{equation}
Here, the spot size $\sigma(x) = \sigma_0 \sqrt{1+(x/x_R)^2}$, where $\sigma_0$ is the beam waist, $x_R = \pi \sigma_0^2 / \lambda_i$ is the Rayleigh range and $\lambda_i$ is the wavelength of the optical tweezer, and $I^{(i)}_0 = 2P^{(i)}$/$\pi \sigma_0^2$  is the peak intensity of the optical tweezer of power $P^{(i)}$ focused onto ion $i$.

Using Eq.~(\ref{eqn:generalOpticalPotential}), one can define the optical trapping frequencies $\omega^{(\alpha)\ \mathrm{opt}}_{i}$ at the focus of the tweezer, i.e. the equilibrium position of ion $i$, in the $\alpha \in \{x,y,z\}$ direction as follows: 
\begin{align}
\omega^{(x) \ \mathrm{opt}}_{i} &=
\sqrt{\frac{\chi_i I_0^{(i)} \lambda_i^2}{2 \pi^2 \sigma_0^4 M_i}}, \label{eqn:generalaxialopttrap}\\
\omega^{(y) \ \mathrm{opt}}_{i} &= \omega^{(z) \ \mathrm{opt}}_{i} =
\sqrt{\frac{\chi_i I_0^{(i)}}{\sigma_0^2 M_i}}, \label{eqn:generaltransverseopttrap}
\end{align}
where,
\begin{equation}
\chi_i = \sum_a \frac{6\pi c^2}{ \omega_{a,i}^3}\left(\frac{\Gamma_{a,i}}{\omega_{a,i} - \omega_{l,i}} + \frac{\Gamma_{a,i}}{\omega_{a,i} + \omega_{l,i}} \right). \label{eqn:generalchi}
\end{equation}

In the case where the optical tweezer is sufficiently far-detuned from all but one excited state, labeled $a^*$, and using the rotating wave approximation, Eq.~(\ref{eqn:generalchi}) can be approximated as,
\begin{equation}
\chi_i \approx \frac{6 \pi c^2 \Gamma_{a^*,i}}{\omega_{a^*,i}^3 \delta_{a^*,i}}, \label{eqn:chi}
\end{equation}
where $\delta_{a^*,i} = \omega_{l,i} - \omega_{a^*,i}$ is the detuning of the optical tweezer from the transition to the excited state $a^*$. 

It is to be noted that the differential AC Stark shift between multiple atomic states will induce a state dependent optical potential and hence trapping frequencies, as seen from Eqs.~(\ref{eqn:generalOpticalPotential}) to (\ref{eqn:generalchi}).
This state dependence can open up opportunities as well as pose complications in certain QIP experiments.
However, in this manuscript, we assume the state dependence to be negligible, and discuss the validity of this assumption in Section \ref{sec:discussion}.

The trap frequencies of the hybrid trap on ion-$i$ $\omega_{\alpha, i}$ would have the following form:
\begin{equation}
    \omega_{\alpha, i} = \sqrt{\left[ \omega^{\mathrm{conv}}_{\alpha} \right]^2 + \left[ \omega^{(\alpha)\ \mathrm{opt}}_{i} \right]^2}.
\end{equation}

When perturbed by an external force, the ions will undergo vibrations that can be written as a linear superposition of collective phonon (or normal) modes.
These phonon modes can be used to mediate effective interactions between internal states of ions for QIP experiments \cite{Cirac1995quantum,Sorensen1999quantum,Zhu2006,kim2009entanglement,Korenblit2012quantum,Wright2017}.
Since $x,y$, and $z$ form the principal axes of the total potential $\phi^{\mathrm{total}}$, the phonon modes do not couple motions along different directions. 
We can obtain the normal mode eigenfrequencies $\mathbf{w}^{(\alpha)}$ and eigenvector matrix $\mathbf{B}^{(\alpha)}$ by diagonalizing the symmetrized mass-weighted Hessian, henceforth referred to as the $\mathbf{A}$-matrices, one for each direction $\alpha$.
The $\mathbf{A}^{(\alpha)}$-matrices have the following form:
\begin{equation}
\mathrm{A}_{ij}^{(\alpha)}
= \frac{1}{\sqrt{M_i}} \left. \left[ \mathrm{Hess}_{\alpha} \left( \phi^{\mathrm{total}} \right) \right]_{ij} \right|_{\{\mathbf{r}^*_i \}_{i=1}^N} \frac{1}{\sqrt{M_j}} \label{eqn:Amatrix}.
\end{equation}
Here, $[ \mathrm{Hess}_{\alpha} (f) ]_{ij} = \partial^2 f / \partial\alpha_i \partial\alpha_j $ denotes the Hessian operation on a function $f$.

We assume that the optical tweezers are placed in such a way that the shift in the equilibrium positions of ions is negligible. 
This is valid for trapping optical tweezers of arbitrary strength and for anti-trapping tweezers that are much weaker than the conventional trap.

It should be noted that the off-diagonal elements of $\mathbf{A}$-matrices are determined solely by the Coulomb potential, while the diagonal elements depend on each term in Eq.~(\ref{eqn:totalpotential}). 
Among the contributions to the diagonal elements of $\mathbf{A}$-matrices, the conventional harmonic trapping potential (Eq.~(\ref{eqn:harmonicpotential})) allows for identical trapping potential for all ions of same mass and charge.
On the other hand, optical tweezers on individual ions gives us control over site-dependent trapping strengths.
Consequently, we also gain limited control over the phonon mode eigenfrequencies and eigenvectors using optical tweezers.
In the perturbative regime, where optical trapping strengths are small and there are negligible changes to the eigenvectors of the system, we obtain the following analytical expression for the normal mode spectrum,

\begin{align}
\begin{split}
\left( \mathrm{w}^{(\alpha)}_m \right)^2
\approx{}& \left(\mathrm{w}^{(\alpha) \ \mathrm{conv}}_m \right)^2 \\
& + \sum_i   \left( \omega^{(\alpha) \ \mathrm{opt}}_{i} \right)^2 \left( \mathrm{B}^{(\alpha) \ \mathrm{conv}}_{im} \right)^2.
\end{split} \label{eqn:linearization}
\end{align}
Here, $\mathbf{w}^{(\alpha) \ \mathrm{conv}}=\left\{ \mathrm{w}^{(\alpha) \ \mathrm{conv}}_m \right\}_m$ is the conventional normal mode spectrum, i.e. for $\phi^{\mathrm{opt}}_{i} = 0$ in Eq.~(\ref{eqn:totalpotential}) for all $i$.
For higher optical trapping strengths from the optical tweezer, the phonon mode frequencies and eigenvectors can be numerically calculated, by diagonalizing the $\mathbf{A}$-matrices in Eq.~(\ref{eqn:Amatrix}).

We demonstrate the control over phonon modes on an $N=5$ ion system with a single optical tweezer shining on ion-$i$, as shown in Fig.~\ref{fig:singletweezer}.
The optical tweezer does not change the $x$-normal modes appreciably, as expected from the weaker confinement along the direction of propagation (Eq.~(\ref{eqn:generalaxialopttrap})).
The normal mode frequencies along $y$ and $z$-directions increase with increasing tweezer strength, unless the $i$-th ion does not take part in that normal mode (e.g. the second highest and lowest modes in Fig.~\ref{fig:singletweezer}(c)).
In the limit $\omega^{(y) \ \mathrm{opt}}_i \approx \omega^{\mathrm{conv}}_y$, the $i$-th ion effectively decouples from rest of the system and the highest mode increases linearly with $\omega^{(y) \ \mathrm{opt}}_i$.

\section{Control of normal modes using optical tweezers} \label{sec:ControlwithTweezers}

\begin{figure*}[t]
    \centering
    \includegraphics[width=\textwidth]{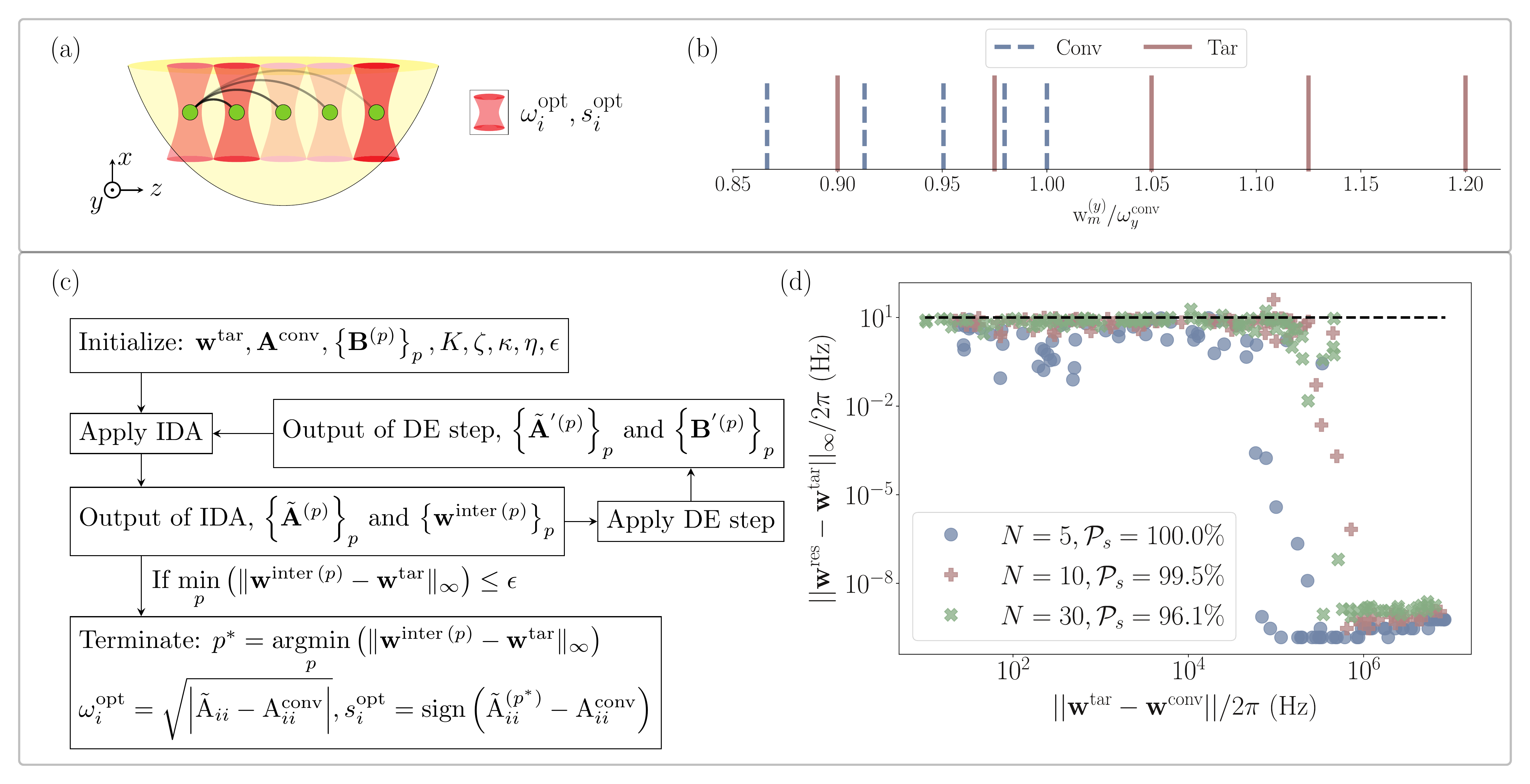}
    \caption{\textbf{The IDADE protocol for finding optical tweezer parameters to obtain $\mathbf{w}^{\mathrm{tar}}$.}
    (a) Schematic of an example system of $N=5$ $^{171}\mathrm{Yb}^+$ ions, with optical tweezers of controllable strength $\omega^{\mathrm{opt}}_i$ and sign $s^{\mathrm{opt}}_i$ (trapping or anti-trapping).
    (b) An example of a conventional normal mode spectrum $\mathbf{w}^{\mathrm{conv}}$ and a target spectrum  $\mathbf{w}^{\mathrm{tar}}$ in the $y$-direction. 
    (c) Schematic of the IDADE protocol to obtain $\omega^{\mathrm{opt}}_i$ and $s^{\mathrm{opt}}_i$.
    (d) Performance of the IDADE protocol for $N=5,10$ and $30$ ion-systems.
    Starting with 1,000 sets of random values of $\omega^{\mathrm{opt}}_i$ and $s^{\mathrm{opt}}_i$ (only 100 are shown for simplicity) , we generate 1,000 solvable sets of $\mathbf{w}^{\mathrm{tar}}$.
    We then benchmark the IDADE protocol by monitoring its ability to output normal mode frequencies $\mathbf{w}^{\mathrm{res}}$ that are within $\epsilon / 2\pi = 10 \ \mathrm{Hz}$ (dashed black line) to $\mathbf{w}^{\mathrm{tar}}$.
    All the points below the dashed line are considered `successful' and we show the success probability of the protocol $\mathcal{P}_s$.
    We initialize the IDADE protocol with a randomly generated ensemble of initial-guesses of the eigenvector matrix $\{ \mathbf{B}^{(p)} \}_p$.
    We empirically find that for $N=5,10$ ions, the required number of IDA iteration is $K=100$, and for $N=30$, $K=600$.
    The IDADE mutation parameters (Eqs.~(\ref{eqn:mutation}) and (\ref{eqn:crossover})) are the same for various $N$, i.e. $\zeta=0.9$, $\kappa=0.5$, $\eta= 0.5$, $p \in \{1,2,...,100\}$).}
    \label{fig:IDADE}
\end{figure*}

In this section, we develop a protocol to solve the inverse problem of finding optical tweezer strengths that are required to achieve a set of eigenfrequencies $\mathbf{w}^{\mathrm{tar}}$ or eigenvectors $\mathbf{B}^{\mathrm{tar}}$ of a target normal mode structure.
Fig~\ref{fig:IDADE}(a) illustrates an example system of five ions with individual optical tweezers of controllable strength.
We focus on a specific direction $\alpha$ without the loss of generality and hence drop the $\alpha$ index on variables.
However, as optical tweezers afford us only limited control over the elements of $\mathbf{A}$-matrices, a solution for a target normal mode structure is not guaranteed, and when a solution exists it may not be unique.

In finding $\mathbf{w}^{\mathrm{tar}}$ and $\mathbf{B}^{\mathrm{tar}}$, we constrain the off-diagonal elements of $\mathbf{A}$-matrices to be equal to those of $\mathbf{A}^{\mathrm{conv}}$-matrices.
For simplicity, we assume that there are no cross-talk between tweezers on neighboring ions, which can be satisfied with precise optical engineering for typical inter-ion spacing of a few micrometers \cite{Monroe2020programmable}.

Our protocol to engineer $\mathbf{w}^{\mathrm{tar}}$ (an example shown in Fig~\ref{fig:IDADE}(b)) closely follows iterative Fourier transform algorithms \cite{Gerchberg1972practical}, except we substitute the Fourier-transform operation with diagonalization.
Henceforth, we call our protocol the iterative diagonalization algorithm (IDA). 
IDA relies on the fact that, for a symmetric matrix (such as $\mathbf{A}$-matrices), diagonalization is a unitary transformation, similar to Fourier transformation.
IDA takes the following parameters as inputs: the target normal mode spectrum $\mathbf{w}^{\mathrm{tar}}$, an initial-guess eigenvector matrix $\mathbf{B}$, $\mathbf{A}^{\mathrm{conv}}$-matrix (for off-diagonal constraints on the $\mathbf{A}$-matrix) and a user-defined accuracy $\epsilon$. 
IDA proceeds in the following order:
(1) We first construct an intermediate matrix $\bar{\mathbf{A}}$ as follows,
\begin{equation}
\bar{\mathrm{A}}_{ij} = \underset{m}{\sum} \mathrm{B}_{im} \left( \mathrm{w}^\mathrm{tar}_m \right)^2 \mathrm{B}^{\intercal}_{mj}. \label{eqn:constructA}
\end{equation}
By construction, $\bar{\mathbf{A}}$ has $\mathbf{w}^{\mathrm{tar}}$ as eigenfrequencies.
However, it may not always satisfy the equality constraints on the off-diagonal elements of the $\mathbf{A}$-matrix. 
Therefore, to enforce the equality constraints, (2) we apply the transformation $T$,
\begin{equation}
\tilde{\mathrm{A}}_{ij} = \left[ T(\bar{\mathbf{A}}) \right]_{ij} =
\begin{cases}
    \bar{\mathrm{A}}_{ii}, \text{ if } i = j, \\
    \mathrm{A}^{\mathrm{conv}}_{ij}, \text{ otherwise}.
\end{cases} 
\end{equation}
(3) We diagonalize $\tilde{\mathbf{A}}$ to obtain the intermittent normal mode eigenfrequencies and eigenvector matrix, $\mathbf{w}^{\mathrm{inter}}$ and $\mathbf{B}^{\mathrm{inter}}$ respectively.
Note that, while $\tilde{\mathbf{A}}$ always satisfies the off-diagonal constraints, $\mathbf{w}^{\mathrm{inter}}$ is not guaranteed to be the same as $\mathbf{w}^{\mathrm{tar}}$.
(4) We repeat steps 1-3, with the intermittent normal mode eigenvector matrix $\mathbf{B}^{\mathrm{inter}}$ as the new initial-guess.

We terminate IDA when either of the following conditions is satisfied:
i) the intermittent normal mode eigenfrequencies, $\mathbf{w}^{\mathrm{inter}}$, agrees with the target normal mode eigenfrequencies, $\mathbf{w}^{\mathrm{tar}}$, within some desired precision $\epsilon$, i.e. $\left\lVert \mathbf{w}^{\mathrm{tar}} - \mathbf{w}^{\mathrm{inter}} \right\rVert_{\infty} \leq \epsilon$ or ii) $\mathbf{w}^{\mathrm{inter}}$ is stuck at an undesired fixed-point $\mathbf{w}^{*}$. 

When the IDA converges (case i), the required parameters for the optical tweezers can be calculated using the following equations,
\begin{align}
\omega^{\mathrm{opt}}_i &=
\sqrt{\left| \tilde{\mathrm{A}}_{ii} - \mathrm{A}^{\mathrm{conv}}_{ii} \right|} \label{eqn:opttrapstrength}, \\
s^{\mathrm{opt}}_i &=
\mathrm{sign} \left( \tilde{\mathrm{A}}_{ii} - \mathrm{A}^{\mathrm{conv}}_{ii} \right) \label{eqn:opttrapsign},
\end{align}
where $\omega^{\mathrm{opt}}_i$ is the required trap frequency of the $i$-th optical tweezer and $s^{\mathrm{opt}}_i=\pm 1$ indicates whether the optical potential should be trapping or anti-trapping, respectively.
For the existence of a fixed-point (case ii), the eigenvectors $\mathbf{B}^{\mathrm{inter}}$ have to be invariant under the transformation $T$, which implies the following,
\begin{equation}
    \sum_{m} \mathrm{B}^{\mathrm{inter}}_{im} \delta \mathrm{w}_m \mathrm{B}^{\mathrm{inter} \ \intercal}_{mj} = 0, \ \forall \ i = j.
\end{equation}
Here, $\delta \mathbf{w} = \mathbf{w}^{*} - \mathbf{w}^{\mathrm{tar}}$.
This further implies $\mathrm{det}(\mathbf{B} \circ \mathbf{B})$ = 0, where $\circ$ is the element-wise (Hadamard) product.

To avoid termination of the IDA at the fixed-points, we modify the protocol to incorporate differential evolution (DE) \cite{Storn1995differential}, a technique for global optimization.
The resulting protocol, referred to as iterative diagonalization algorithm with differential evolution (IDADE), is shown schematically in Fig.~\ref{fig:IDADE}(c).
The IDADE protocol functions as follows:
(1) We first begin with an ensemble of initial-guesses for the eigenvector matrix, indexed by population index $p$, $\{\mathbf{B}^{(p)}\}_p$.
(2) Then, we apply IDA in parallel for each initial-guess, until at least one instance converges, or we reach a maximum user-defined iteration number $K$.
(3) In case of no convergence, DE is applied (see below) to the output $\{\tilde{\mathbf{A}}^{(p)}\}_p$ matrices (at the $K$-th iteration) of IDA.
(4) The resultant matrices are diagonalized to obtain eigenvector matrices, $\{\mathbf{B}'^{(p)}\}_p$, which will be used as an updated ensemble of initial-guesses for IDA again.

A DE step mutates the ensemble of $\tilde{\mathbf{A}}$-matrices in the following way: 
(1) $\{\tilde{\mathbf{A}}^{(p)}\}_p$ are diagonalized to obtain eigenfrequencies $\{\mathbf{w}^{(p)}\}_p$.
(2) We determine the index $p^*$ as follows
\begin{equation}
    p^* = \underset{p}{\mathrm{argmin}} \left( \left \lVert \mathbf{w}^{(p)} - \mathbf{w}^{\mathrm{tar}} \right \lVert_\infty \right).
\end{equation}
(3) We define $\mathbf{x}^{(p)} = \mathrm{diag}\left(\mathbf{\tilde{A}}^{(p)}\right)$. The ensemble of $\{\mathbf{x}^{(p)}\}_p$ is mutated in the following way for each index $p$:
\begin{align}
\begin{split}
u \left(\mathbf{x}^{(p)}\right)
={}& \mathbf{x}^{(p)} + \zeta \left( \mathbf{x}^{(p^*)}-\mathbf{x}^{(p)} \right) \\
& + \kappa \left( \mathbf{x}^{(q)} - \mathbf{x}^{(r)} \right)
\end{split} \label{eqn:mutation}
\end{align}
with $\{p,q,r\}$ mutually different, and $q$ and $r$ chosen randomly.
Intuitively, the parameter $\zeta$ ``biases'' the ensemble towards the best performing entity ($p^*$), and $\kappa$ ``randomizes'' the ensemble to explore the entire solution space.
We keep the mutated value of each element of $\mathbf{x}^{(p)}$ with a probability $\eta$, and proceed with $\mathbf{x}'^{(p)}$ given by, 
\begin{equation}
\mathrm{x}'^{(p)}_i =
\begin{cases}
    \left[ u \left( \mathbf{x}^{(p)} \right) \right]_i, \text{ with probability } \eta, \\
    \mathrm{x}^{(p)}_i, \text{ otherwise. }
\end{cases} \label{eqn:crossover}
\end{equation}
In Eqs.~(\ref{eqn:mutation}) and (\ref{eqn:crossover}) $\zeta, \kappa$ and $\eta$ are empirically chosen parameters of DE.
(4) Using $\mathbf{x}'^{(p)}$, we construct $\tilde{\mathbf{A}}'^{(p)}$ and diagonalize it to obtain, $\mathbf{w}'^{(p)}$.
For each index $p$, depending on whether $\mathbf{w}'^{(p)}$ or $\mathbf{w}^{(p)}$ is closer to $\mathbf{w}^{\mathrm{tar}}$, we keep either $\tilde{\mathbf{A}}'^{(p)}$ or $\tilde{\mathbf{A}}^{(p)}$, respectively, for the next round of IDA.

In Fig.~\ref{fig:IDADE}(d), we show the performance of the IDADE protocol in engineering the normal mode spectrum of $N=5, 10$ and $30$ systems.
We are able to reach a wide range of target spectra, within a tolerance of $\epsilon= 2\pi \times 10 \ \mathrm{Hz}$.
Our choice of $\epsilon / 2\pi = 10 \ \mathrm{Hz}$ is motivated by practical experimental limits in stabilizing the normal mode frequencies \cite{johnson2016active}.
When the target mode frequencies are much larger than the conventional mode frequencies, the problem becomes trivial as the Coulomb interactions are negligible compared to the optical potential.
We find that in this case, the accuracy is increased significantly to $\epsilon / 2\pi  \approx 10^{-8}$ Hz.
The success probability of the IDADE protocol decreases with increasing system size, which can potentially be mitigated by increasing the number of iterations $K$ as well as optimizing the DE parameters.

Solving for a target set of normal mode eigenvectors $\mathbf{B}^{\mathrm{tar}}$ is equivalent to finding the optical tweezer parameters in the $\mathrm{A}$-matrix, such that 
\begin{equation}
    \sum_{i,j} \mathrm{B}^{\mathrm{tar} \ \intercal}_{mi} \mathrm{A}_{ij} \mathrm{B}^{\mathrm{tar}}_{jn} = \mathrm{D}_{mn},
    \label{eqn:diagonalization}
\end{equation}
where $\mathbf{D}$ is an arbitrary $N \times N$ real diagonal matrix.
The problem can be formulated in terms of two sets of linear equations.
First, we determine a set of basis matrices, indexed by $k$, $ \{ \hat{\mathbf{A}}^{(k)} \}_k$, that spans the space of all $N \times N$ real matrices with eigenvectors $\mathbf{B}^{\mathrm{tar}}$, by solving
\begin{equation}
    \sum_{i,j} \mathrm{C}_{ijmn}\mathrm{A}_{ij} = 0, \quad \forall \ m \neq n,
    \label{eqn:eigvecbasis}
\end{equation}
where $\mathrm{C}_{ijmn} = \mathrm{B}^{\mathrm{tar}}_{im} \mathrm{B}^{\mathrm{tar}}_{jn}$.
Next, we search for a set of real numbers $\{ \gamma_k \}_k$ for which $ \mathrm{A}_{ij} = \sum_{k} \gamma_k \hat{\mathrm{A}}_{ij}^{(k)}$ satisfies the off-diagonal Coulomb constraints, i.e.
\begin{equation}
    \sum_{k} \gamma_k \hat{\mathrm{A}}_{ij}^{(k)} = \mathrm{A}^{\mathrm{conv}}_{ij}, \quad \forall \ i \neq j. \label{eqn:eigveccoeff}
\end{equation}
We can find the desired optical tweezer parameters from $\mathbf{A}$ as per Eqs.~(\ref{eqn:opttrapstrength}) and (\ref{eqn:opttrapsign}), when a solution for Eq.~(\ref{eqn:eigveccoeff}) exists.

\section{Application of normal mode control}
\label{sec:application}

The control over phonon mode eigenfrequencies and eigenvectors afforded by optical tweezers would allow for exploring fundamental physics problems as well as solving bottlenecks of some QIP experiments.
Optical tweezer-induced control of local trap frequencies can be used in simulating quantum thermodynamical properties of a multi-species system of different masses, within an experimentally simpler system of a single ion-species.
In this section, we describe the building block for this kind of a simulator, to extract fundamental thermodynamic properties of one ion ($^{133}$Ba$^+$ for example) using another species ($^{171}$Yb$^+$) and an optical tweezer.
We also describe how controlling the eigenvectors potentially solves an important limitation of multi-species quantum information protocols, i.e. weak coupling between species with a large mass-imbalance \cite{sosnova2020character}.

\subsection{Simulating quantum thermodynamics of a species with programmable effective mass} \label{sec:Thermodynamics}

\begin{figure*}[t]
\centering
\includegraphics[width=\textwidth]{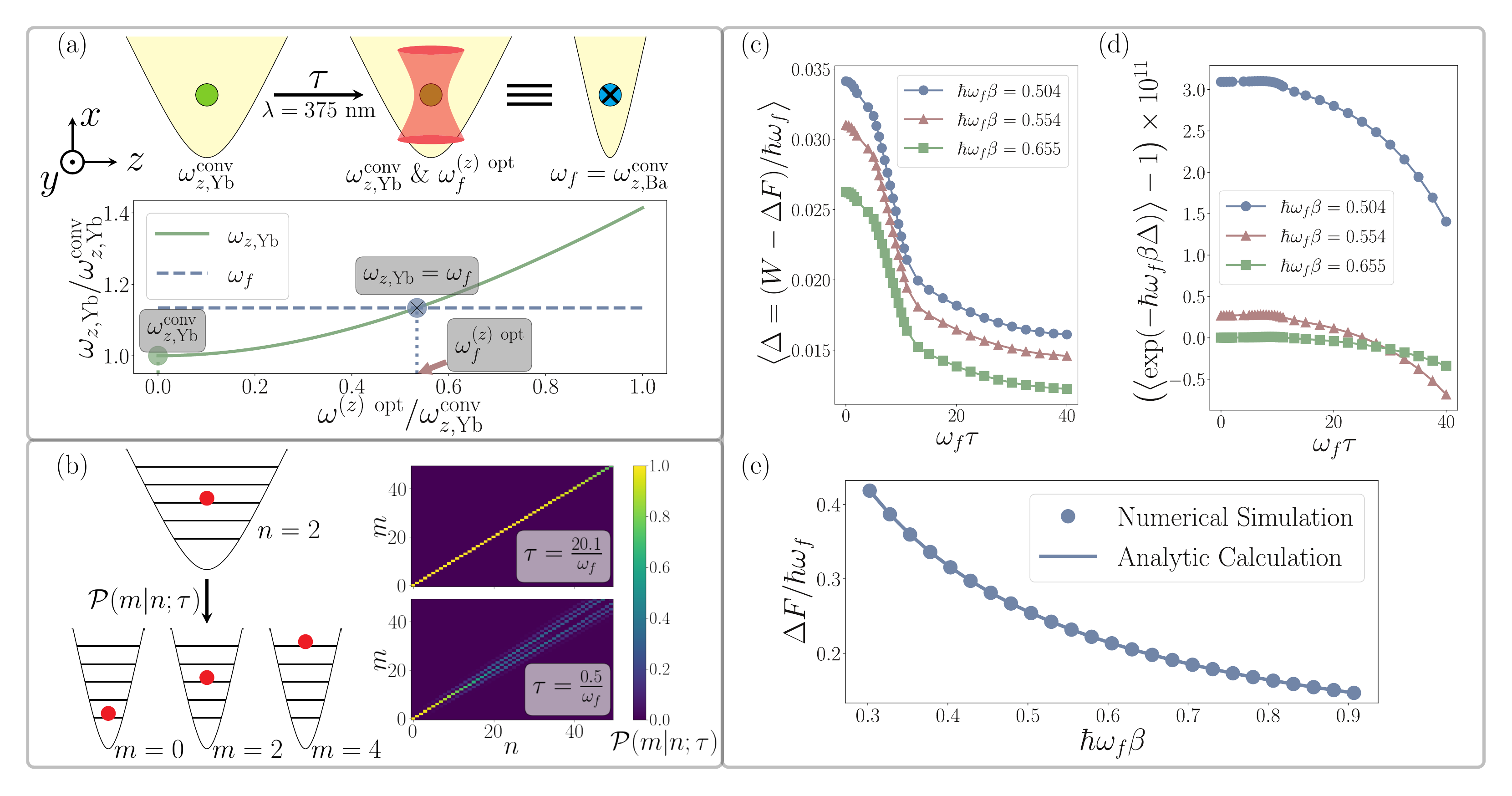}
\caption{\textbf{Determination of thermal free energy of a different ion species ($^{133}\mathrm{Ba}^{+}$ in this case) using optical tweezer on a single $^{171}\mathrm{Yb}^{+}$.}
The normal mode in the axial-direction (labelled as the z-direction) is used to explicate the protocol. 
(a) Description of the experimental scheme (see text).
The axial trap frequency $\omega_{z,\mathrm{Yb}}$ of the $^{171}\mathrm{Yb}^{+}$ ion increases as a function of the optical tweezer strength, characterized by the optical trapping frequency $\omega^{(z) \ \mathrm{opt}}$ (Eq.~\ref{eqn:generalaxialopttrap}).
At $\omega_{z,\mathrm{Yb}} = \omega_{f}$, the trap frequency becomes identical with that of a $^{133}\mathrm{Ba}^{+}$ ion in the conventional trap.
(b) The conditional transition probabilities P(m$\vert$n, $\tau$) for transitioning to the Fock state $|m\rangle$ in the final Hamiltonian from $|n\rangle$ in the initial Hamiltonian are plotted, for two switching rates (${1}/\tau$).
(c) The dependence of the first moment of the work probability distribution (See Appendix~\ref{apdx:workdist}) versus switching times for various temperatures. 
(d) The first moment of the Jarynski's function \cite{jarzynski1997nonequilibrium,Talkner2007}, showing insensitivity to switching times and thus giving us a way to extract the thermal free energy (with an error $\sim 10^{-11}$) from fast diabatic processes.
(e) The thermal free energy difference of the axial mode of $^{133}\mathrm{Ba}^{+}$ ion, with respect to the axial mode of $^{133}\mathrm{Yb}^{+}$ ion, as a function of temperature (see text). 
The points indicate the values numerically simulated for the proposed scheme and the solid line shows the analytically obtained result, illustrating a good match.}
\label{fig:JE_panel}
\end{figure*}

Quantum thermodynamics of few-body systems empowers us to unravel the origin of irreversibility in a bottom-up way \cite{Strasberg2017}, design devices for heatronics and thermal transport \cite{Yang2013, doi:10.1063/1.4905132, PhysRevE.89.062109,doi:10.1063/1.4941405, doi:10.1063/1.5001072}, construct measurement-driven thermal engines \cite{Elouard2017,PhysRevE.96.022108}, and formulate resource-theoretic description of thermodynamics \cite{RevModPhys.91.025001,ThomasGuffNathanA.McMahonYuvalR.Sanders2019}.
Recent experiments with trapped ions probed phonon-counting statistics in a dynamical quantum system and demonstrated the validity of important quantum thermodynamic theorems such as Jarzynski's equality \cite{jarzynski1997equilibrium,jarzynski1997nonequilibrium,Talkner2007,PhysRevE.86.011111,Hartmann2019}.
Here, we propose the use of Jarzynski's equality to extract thermal free energy of a system of programmable effective mass, by varying the optical tweezer strength in the simulator.
Free energy, which is an equilibrium property of a thermodynamic system, plays a central role in statistical mechanics \cite{LeBellac2004} from which other characteristic features of the system can be derived.
Jarzynski's equality gives a practical way to measure changes in the thermal free energy of a system from far-from-equilibrium dynamics, e.g. by quickly changing the optical tweezer parameters.

The protocol proceeds as follows. 
(1) We start by preparing an ion of mass $M_{\mathrm{expt}}$ in a thermal state with inverse-temperature $\beta$ in a conventional trap.
(2) We do a phonon-number-resolved measurement \cite{An2015}, which projects the state onto a Fock state $\ket{n}$ with energy $E_n$.
Here, $\ket{n}$ is an eigenstate of the initial Hamiltonian (conventional trap).
(3) We ramp the optical tweezer strength from zero to a pre-calibrated value in time $\tau$.
The final tweezer strength corresponds to a value at which the trapping frequency $\omega_{f}$ of the experimental ion of mass $M_{\mathrm{expt}}$ is the same as that of another species of mass $M_{\mathrm{tar}}$ in a conventional trap.
(4) We perform another phonon-number-resolved measurement, which projects the state into a Fock state $\ket{m}$ with energy $E'_m$.
Here, $\ket{m}$ is an eigenstate of the final Hamiltonian (conventional trap plus the optical tweezer).
Assuming that the evolution is conservative, the work $W$ performed on the system will be $E'_m - E_n$.
For a finite $\tau$, the evolution in previous steps would be out-of-equilibrium, which will result in shot-to-shot fluctuations in $W$, even when the starting Fock state is the same, i.e. $m$ is not necessarily the same as $n$.
Further, there would be additional fluctuations in $W$ arising from the sampling of the initial Fock state $\ket{n}$ from the thermal distribution.
These fluctuations result in a distribution $\mathcal{P}(W)$ of work values.
Jarzynski's equality\cite{jarzynski1997nonequilibrium,Talkner2007} connects the averaged (over $\mathcal{P}(W)$) exponentiated work to the thermal free energy change $\Delta F$ of the system, as follows,
\begin{equation}
    \langle \exp(-\beta (W - \Delta F)) \rangle = 1. \label{eqn:Jazynskiequality}
\end{equation}
From Eq.~(\ref{eqn:Jazynskiequality}) , we can compute the free energy of the ion of mass $M_{\mathrm{tar}}$ with respect to the initial (known or referenced) free energy of mass $M_{\mathrm{expt}}$.

For an illustration of the above mentioned protocol, we specifically use $M_{\mathrm{expt}}= M_{\mathrm{Yb}}$ = 171 a.m.u. corresponding to a single $^{171}\mathrm{Yb}^{+}$ ion in a conventional trap.
The target species is $^{133}\mathrm{Ba}^{+}$ ($M_{\mathrm{tar}} = M_{\mathrm{Ba}}$ = 133 a.m.u.) whose thermal free energy at inverse temperature $\beta$ would be determined from the protocol, by ramping the power of the optical tweezer linearly.
As shown in Fig. ~\ref{fig:JE_panel}(a), the axial trap frequency $\omega_{z,\mathrm{Yb}}$ of $^{171}\mathrm{Yb}^+$ in the hybrid trap matches the $^{133}\mathrm{Ba}^{+}$ axial trap frequency in the conventional trap, $\omega_{f} = \omega^{\mathrm{conv}}_{z,\mathrm{Ba}} = \omega^{\mathrm{conv}}_{z,\mathrm{Yb}} \sqrt{M_{\mathrm{Yb}}/M_{\mathrm{Ba}}}$ at time $\tau$.

\begin{figure*}[t]
    \centering
    \includegraphics[width=0.8\textwidth]{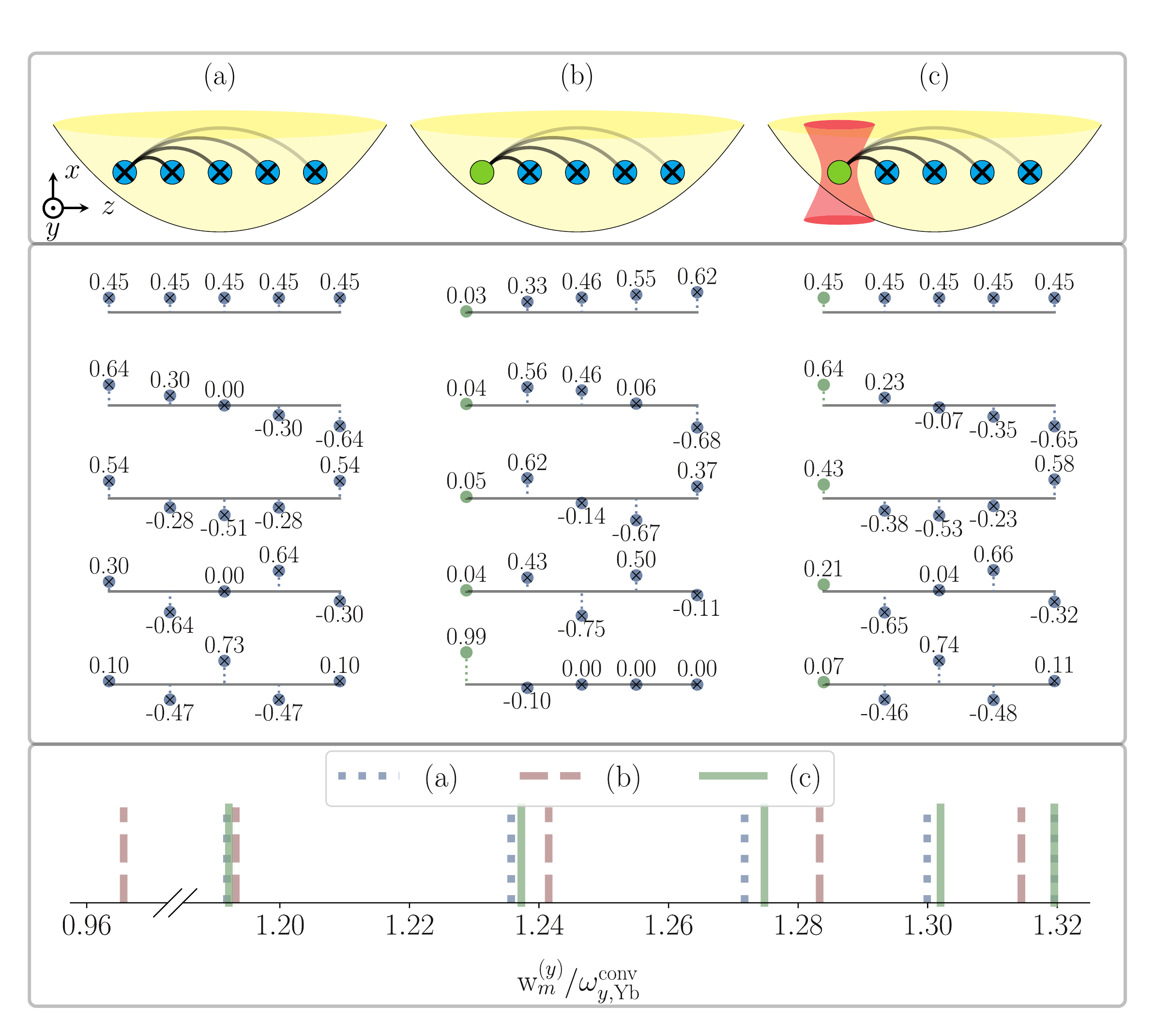}
    \caption{\textbf{`Correcting' the normal mode eigenvectors in a dual-species system.}
    (Top panel) Schematic of (a) $^{133}\mathrm{Ba}^+$ ions (blue node with cross) and (b) a dual-species system of $^{171}\mathrm{Yb}^+$ (green node) and $^{133}\mathrm{Ba}^+$ ions in a conventional trap.
    (c) the same dual-species system with an optical tweezer applied on $^{171}\mathrm{Yb}^+$.
    We consider the conventional trap frequencies for $^{171}\mathrm{Yb}^+$ are $\omega^{\mathrm{conv}}_{x,\mathrm{Yb}} = 2 \pi \times 1.2 \ \mathrm{MHz}, \ \omega^{\mathrm{conv}}_{y,\mathrm{Yb}} = 2 \pi \times 1 \ \mathrm{MHz}$ and $\omega^{\mathrm{conv}}_{z,\mathrm{Yb}} = 2\pi \times 0.2 \ \mathrm{MHz}$.
    % Therefore, the conventional trap frequencies for $^{133}\mathrm{Ba}^+$ are $\omega^{\mathrm{conv}}_{x, \mathrm{Ba}}/2\pi \approx \omega^{\mathrm{conv}}_{y, \mathrm{Ba}} =2\pi \times 1.32 \ \mathrm{MHz}$ and $\omega^{\mathrm{conv}}_{z, \mathrm{Ba}}= 2\pi \times  0.23 \ \mathrm{MHz}$ (see Appendix~\ref{apdx:convtrap}).
    (Middle panel) Corresponding $y$-direction ion-motion eigenvectors $\bar{\mathbf{B}}^{(y)}$ of normal modes ranked in descending order of their frequencies from the top.
    The tweezer in (c) has a $y$-trap strength of $0.861\ \omega^{\mathrm{conv}}_{y,\mathrm{Yb}}$, which makes the ion-motion eigenvector of the highest frequency normal mode same as that in configuration (a).
    (Bottom panel) Corresponding $y$-direction eigenfrequencies.
    }
    \label{fig:multispecies}
\end{figure*}

As seen from the conditional transition probability $\mathcal{P}(m \vert n, \tau)$ for all ($m,n$) pairs in Fig.~\ref{fig:JE_panel}(b), large switching time (${\tau} \gg 1 / \omega_{f}$) corresponds to the adiabatic limit \cite{Band1992,PhysRevA.72.012114}, $n=m$, while faster switching time ($ \tau \lesssim 1 / \omega_{f}$) leads to cross-excitations of $m \ne n$ phonon states. 
It should be noted that cross-excitations of the kind $m+n \ne 2k \ \forall \ k \in \mathbb{Z}^+$ are disallowed due to conservation of parity symmetry in the Hamiltonian.
Also, for higher $n$, cross-excitations are more probable due to enhanced overlap between real-space wavefunctions for larger $(m,n)$ pairs.
The work distribution $\mathcal{P}(W)$ is derived from $\mathcal{P}(m \vert n, \tau)$ as follows, \cite{Talkner2007,PhysRevLett.101.070403} 
\begin{equation}
    \mathcal{P}(W) = \sum_{m,n} \mathcal{P}(m\vert n) \mathcal{P}(n, \beta)\:\: \delta \left(W - \left(\frac{E'_m - E_n}{\hbar \omega_{f}} \right) \right), \label{pW_defn_eqn}
\end{equation}
where $\mathcal{P}(n, \beta)$ is the probability of sampling a Fock state from the the initial thermal distribution (see Appendix~\ref{apdx:workdist} for the detailed work distribution for this example).
As seen in Fig.~\ref{fig:JE_panel}(c), the dimensionless averaged work relative to the thermal free energy change $\langle \Delta \rangle$ varies with both the dimensionless switching time $\omega_{f} \tau$ and the temperature of the initial distribution.
Only in the adiabatic limit, the averaged work done approaches the thermal free energy change.
From Fig.~\ref{fig:JE_panel}(d), we see that the error associated with the averaged exponentiated work $\langle \exp(-\beta \hbar \omega_{f} \Delta)\rangle$ agrees with Jarzynski's equality Eq.~(\ref{eqn:Jazynskiequality}) at $(10^{-11})$ level for various switching time and temperature.
This low error leads to an accurate estimation (at $10^{-11}$ level) of the thermal free energy change ($\Delta F \approx -\ln{\langle \exp(-\beta W) \rangle} / \beta$) from non-equilibrium experiments at short switching times, as seen in the comparison with the analytically calculated (from the partition function) thermal free energy change \cite{Sajjan2020experimental} in Fig.~\ref{fig:JE_panel}(e).
Note that the ability to ramp the optical tweezer fast allows us to avoid adverse effects such as motional heating from fluctuating electric fields \cite{PhysRevA.61.063418}, drifts and fluctuations in the conventional trap and optical tweezer parameters, and possibly from spontaneous emission effects.

\subsection{Enhancing normal mode couplings in a mass-imbalanced system}\label{sec:Multispecies}

Quantum gates between ions of different elements offer advantages to combine positive aspects of the species - such as superior quantum memory with longer wavelength optical readout amenable to matured commercial optical technologies \cite{inlek2017multispecies}.
Another significant advantage of using a multi-species ion system is the ability to perform sympathetic cooling on the entire system without destroying the coherence of the computational ions \cite{sosnova2020character}.
Sympathetic cooling can be used to limit unwanted heating in a quantum processor, allowing longer quantum experiments.
However, a major challenge of working with a multi-species system, especially when there is a large mass-imbalance between the ion-species, is that the motion of the two species can decouple from each other \cite{sosnova2020character}.
Therefore, both phonon-mediated quantum gates and sympathetic cooling of the entire ion system take longer and may even be impractical. 
This problem can potentially be remedied by modifying the trapping potential using optical tweezers on the ions.
We can either `correct' the eigenvectors, such that the tweezer-modified modes better resemble the conventional modes for a single-species system, or we can possibly engineer optimized mode structures for efficient multi-species gates and sympathetic cooling.

Note that for a multi-species system, the eigenvectors obtained from the diagonalization of the $\mathbf{A}$-matrix defined in Eq.~(\ref{eqn:Amatrix}) contains a mass-weighting.
Therefore, in the case of a multi-species system, the physical displacement of the ions is given by mass-weighted eigenvector matrices $\bar{\mathbf{B}}^{(\alpha)}$ defined by,
\begin{equation}
    \bar{\mathrm{B}}^{(\alpha)}_{im} = \frac{1}{\sqrt{M_i}} \mathrm{B}^{(\alpha)}_{im} \ \text{ normalized along index $i$,}  \label{eqn:ionmotionevecs}
\end{equation}
where the normalization ensures that $\sum_{i} \bar{\mathrm{B}}^{(\alpha)}_{im} \bar{\mathrm{B}}^{(\alpha)}_{im} = 1$ for all $m$ and $\alpha$.

To illustrate the use of tweezers to `correct' for the mass-imbalance, we consider the normal modes along a spatial direction ($y$) of a multi-species trapped ion system composed of a single $^{171}\mathrm{Yb}^+$ ion and four $^{133}\mathrm{Ba}^+$ ions.
Fig.~\ref{fig:multispecies} schematically describes the system.
A comparison between Fig.~\ref{fig:multispecies}(a) for all $^{133}\mathrm{Ba}^+$ ions and Fig.~\ref{fig:multispecies}(b) for the mixed-species system shows the decoupling of $^{171}\mathrm{Yb}^+$ ions from the rest of the system.
By applying an optical tweezer on the $^{171}\mathrm{Yb}^+$ ion (Fig.~\ref{fig:multispecies}(c)), the eigenvectors of the highest mode (the center-of-mass mode) of the conventional single-species system is restored.
However, the eigenfrequencies of the system have also been modified (bottom panel of Fig.~\ref{fig:multispecies}).
Here, the desired optical tweezer strength is obtained by matching the trap frequency $ \omega_{y,\mathrm{Yb}}$ of $^{171}\mathrm{Yb}^+$ to the conventional trap frequency $\omega^{\mathrm{conv}}_{y,\mathrm{Ba}}$ of $^{133}\mathrm{Ba}^+$ .
The protocols described in Section~\ref{sec:ControlwithTweezers} can be followed to find the desired tweezer parameters for a system requiring multiple optical tweezers.

Equation~(\ref{eqn:ionmotionevecs}) implies that $\bar{\mathbf{B}}^{(\alpha)}$ is not unitary for a multi-species system.
Hence, it is not possible to simultaneously match all the eigenvectors of the multi-species system to that of the conventional single-species system. 

% ========
% End Body
% ========

% ================
% Begin Discussion
% ================
\section{Discussions}\label{sec:discussion}
In this work, we presented a scheme where the phonon modes of a conventional trapped ion system can be modified using AC Stark effect from an array of optical tweezers.
We demonstrated an algorithm for determining the required tweezer strengths to obtain a target set of phonon mode frequencies or eigenvectors.
Using the tweezer-mediated-control of phonon modes, we can build a simulator for investigating quantum thermodynamics of multi-species systems.
Tweezers can also be used to prevent decoupling of the motion of different species due to mass-imbalance, for example, in multi-species QIP experiments for fast quantum gates and efficient cooling \cite{inlek2017multispecies,sosnova2020character}.
In order to change the normal mode frequencies or eigenvectors of a conventional trap significantly, the trap frequencies due to the tweezer potentials have to be comparable to the conventional trap frequency.
The strength of the tweezer potential can be increased by reducing its detuning $\delta$ or the beam waist $\sigma_0$, or by increasing its peak optical intensity $I_0$, as seen from  Eqs.~(\ref{eqn:generaltransverseopttrap}) and (\ref{eqn:generalchi}).
However, each of these actions can lead to adverse effects in experiments, especially when dealing with qubit-states for QIP.
Reducing the detuning or increasing the peak intensity will increase the spontaneous emission rate as well as differential AC Stark shift between qubit states.
These effects can be a source of decoherence in QIP experiments.
Differential AC Stark shift also makes the trap frequency dependent on the qubit or spin states, which could pose challenges as well as opportunities to explore spin-phonon interaction physics.
Obtaining higher power and hence intensity can especially be challenging for ultra-violet wavelength lasers.
The achievable beam waist is limited by the resolution of the optical system.
The temperature of the ion determines its spatial localization and therefore poses a fundamental lower limit to usable beam waist \cite{cetina2020quantum}.

\begin{table}[ht]
\centering
\caption{Parameters of the optical trapping with tweezers.}
\begin{tabular}{cccc}
\hline
\hline
Wavelength (nm) & 375 & 532 & 1064 \\
Beam waist ($\mu$m) & 0.458 & 0.649 & 1.298 \\
Power (W) & 0.0871 & 2.562 & 61.6 \\
AC Stark shift ($\mathrm{MHz} \cdot h$) & -900 & -1800 & -7000 \\
Differential AC Stark shift ($\mathrm{kHz} \cdot h$) & 778 & 130 & 400 \\
Optical trap frequency ($\mathrm{MHz} \cdot 2\pi$) & 1 &  1 & 1 \\
Differential trap frequency ($\mathrm{kHz} \cdot 2\pi$) & 0.44 & 0.036 & 0.028 \\
Off-resonant scattering rate ($\mathrm{s}^{-1}$) & 7200 & 230 & 70 \\
\hline
\hline
\end{tabular}
\label{tab:optparams}
\end{table}

Table \ref{tab:optparams} provides examples of optical trapping parameters for $^{171}$Yb$^+$ to achieve a 1 MHz optical trap frequency, with $\ket{6^2\mathrm{S}_{1/2}, {\mathrm{F}} = 0, m_{\mathrm{F}}= 0}$ and $\ket{6^2\mathrm{S}_{1/2}, {\mathrm{F}} = 1, m_{\mathrm{F}}= 0}$ providing the qubit states. 
Off-resonant scattering rate is the sum of the scattering rates from all relevant excited states to the ground state $\ket{6^2\mathrm{S}_{1/2}}$ (considering all hyper-fine states of the ground state).  
The strongest relevant atomic transition for the wavelengths shown in Table \ref{tab:optparams} is $\mathrm{S}_{1/2} \rightarrow \mathrm{P}_{1/2}$ at 369.5 nm (see Appendix~\ref{apdx:opttrap}).
As expected, in the far detuned regime, a higher optical power is necessary to attain the same optical trap frequency.
However, the differential AC Stark shift and hence the differential trap frequency, and the off-resonant scattering rate from atomic transitions are significantly lower than that in the near-detuned regime.

Intensity fluctuations in the optical tweezer beams at the ion location, from either laser power fluctuations or beam pointing instabilities, may lead to several adverse effects.
Fluctuations in differential AC Stark shifts could be a source of dephasing in QIP experiments.
Fluctuations in optical trap frequencies could lead to fluctuations in the normal mode frequencies and motional heating of ions.
For example, in order to achieve a target accuracy of $\epsilon/2\pi=10$ Hz of a normal mode frequency at 0.5 MHz (using IDADE algorithm in Section \ref{sec:ControlwithTweezers}), the relative intensity fluctuations in the tweezer beam has to be $10^{-5}$, which could be experimentally challenging.

Our choice of axial modes in Section \ref{sec:Thermodynamics} is motivated by the lower frequency scale and weaker mass-dependence of the axial mode frequency compared to transverse modes.
The lower frequency scale necessitates less optical power and ensures minimal work done by the tweezer on higher frequency transverse modes.
The weak mass dependence of the axial mode frequency allows us to effectively simulate the mechanical properties of systems with large mass differences.
While our specific proposal in this manuscript simulates properties of a single ion of different mass, the local control afforded by optical tweezers allow scaling the system to a large number of ions, with local control over effective mass.

While we considered red-detuned Gaussian beams in this manuscript, one can also use blue-detuned tweezers with an intensity minimum at the ion (such as a Laguerre-Gaussian mode $\mathcal{L}_{0}^{1}$) to minimize some of the adverse effects such as the rate of off-resonant scattering of photons and differential AC Stark shifts.
However, blue-detuned tweezers require more sophisticated optical engineering. 
Further, because of the UV atomic transition in most ions \cite{ozeri2007errors}, the necessity to work with high-power blue-detuned light might be experimentally challenging.
% ==============
% End Discussion
% ==============

% =====================
% Begin Acknowledgments
% =====================

\acknowledgments
We acknowledge discussions with Roger Melko, Roger Luo, Stefanie Czischek, Peter Zoller, Tobias Olsacher, Lukas Sieberer and Chung-You Shih. We acknowledge financial support from 
Canada First Research Excellence Fund (CFREF) through the Tranformative Quantum Technologies (TQT) program, Natural Sciences and Engineering Research
Council of Canada's  Discovery (RGPIN-2018-05250)
program, and Institute for Quantum Computing. RI is also supported by an Early Research Award from the Government of Ontario, and Innovation, Science and Economic Development Canada (ISED). F.R.'s research at Perimeter Institute is supported in part by the Government of Canada through the Department of Innovation, Science and Economic Development Canada and by the Province of Ontario through the Ministry of Economic Development, Job Creation and Trade.

% ===================
% End Acknowledgments
% ===================

% ================
% Begin References
% ================

\bibliography{ref}

% ==============
% End References
% ==============

% ==============
% Begin Appendix
% ==============

\appendix

\section{Conventional trapping}
\label{apdx:convtrap}

In this paper, we consider a Paul (quadrupole) trap \cite{paul1990electromagnetic,wineland1997experimental,matteo2020quantum} consisting of RF and DC electrodes as the conventional trap.
The form of the mass and charge dependent trapping frequencies $\omega^{\mathrm{conv}}_{\alpha}(M_i,q_i)$ (where $\alpha \in \{ x,y,z \}$) generated by the DC and RF electrodes on the ions are as follows:
\begin{align}
\omega^{\mathrm{conv}}_z (M_i,q_i) &= \sqrt{\frac{q_iV_{\mathrm{DC}}\xi_z}{M_i}}, \\
\omega^{\mathrm{conv}}_x (M_i,q_i)
&= \sqrt{-\frac{q_iV_{\mathrm{DC}}\xi_x}{M_i}
+ \frac{q_i^2V_{\mathrm{RF}}^2\psi_x^2}{2M_i^2\Omega_{\mathrm{RF}}^2}}, \\
\omega^{\mathrm{conv}}_y (M_i,q_i)
&= \sqrt{-\frac{q_iV_{\mathrm{DC}}\xi_y}{M_i}
+ \frac{q_i^2V_{\mathrm{RF}}^2\psi_y^2}{2M_i^2\Omega_{\mathrm{RF}}^2}}.
\end{align}

Here, $\xi_{\alpha}, \psi_{\alpha}$ are the trap geometric factors, $V_{\mathrm{DC}},V_{\mathrm{RF}}$ are the DC and RF peak voltages on the respective electrodes and $\Omega_{\mathrm{RF}}$ is the frequency of the RF electrode.

\section{Optical trapping}
\label{apdx:opttrap}

\begin{figure*}[t]
    \centering
    \includegraphics[width=0.8\textwidth]{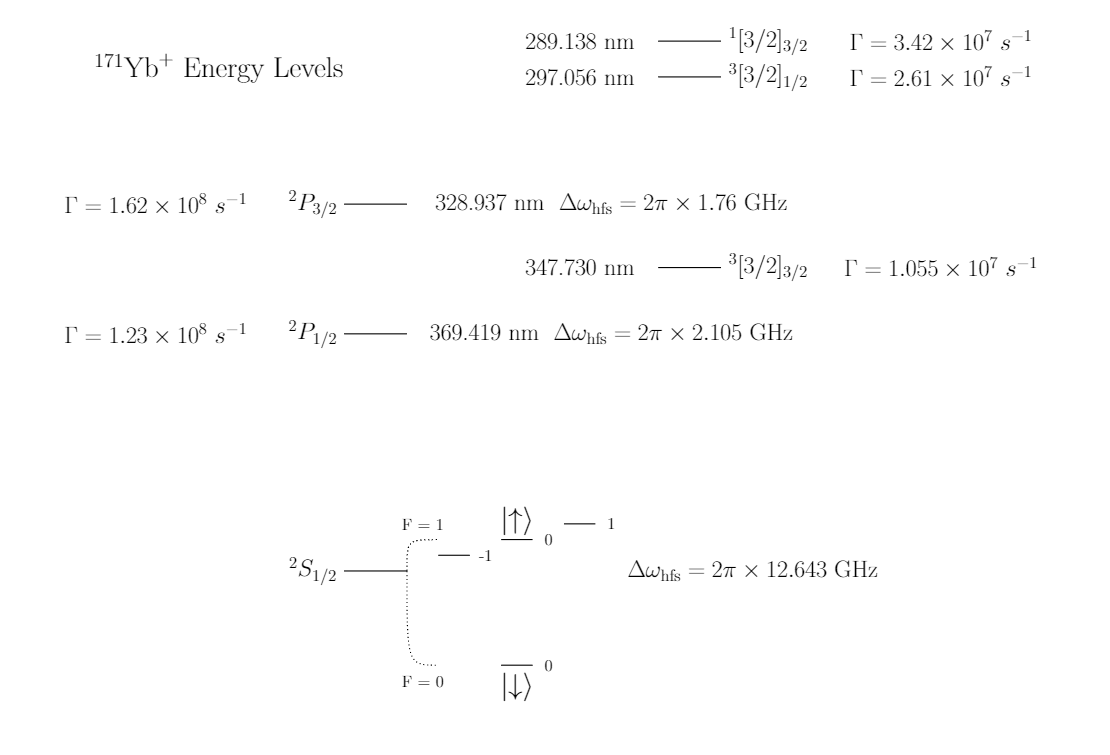}
    \caption{\textbf{Energy levels of $^{171}$Yb$^+$.} Electronic states of $^{171}$Yb$^+$ ion with wavelengths, retrieved from NIST Atomic Spectra Database, Einstein A coefficients \cite{Morton_2000}, denoted as $\Gamma$, and hyper-fine energy splitting \cite{585119, PhysRevA.49.3351, Berends:92, PhysRevA.97.032511}, denoted as $\Delta \omega_{\mathrm{hfs}}$. As an exception, $\Gamma$ of $^3[3/2]_{3/2}$ state (347.730 nm) is not obtained from any references, but derived from $\Gamma$ of $^3[3/2]_{1/2}$ state (297.056 nm) using the Wigner–Eckart theorem. }
    \label{fig:Yblevels}
\end{figure*}

For a two-level system composed of electronic states, at small laser detunings, and using the rotating wave approximation, the AC Stark shift of the ground atomic level for the $i$th ion is given by 
\begin{equation}
    \phi^{\mathrm{opt}}_{s,i}(\mathbf{r}_i) = \frac{\hbar}{2} \delta_{s,i} \left( \sqrt{1 + \frac{\Omega_{s,i}^2}{\delta_{s,i}^2}} - 1 \right), 
\label{A1}
\end{equation}
where $\delta_{s,i}$ is the laser detuning with respect to an excited state $s$, and $\Omega_{s,i}$ is the Rabi frequency of the transition between the ground state and an excited state $s$. Rabi frequency is defined as 
\begin{equation}
     \Omega_{s,i}^2 = \Gamma_{s,i} \frac{6\pi c^2}{\hbar \omega_{s,i}^3} \ I_i(\mathbf{r}_i - \mathbf{r}^*_i), 
\end{equation}
where $\Gamma_{s,i}$ and $\omega_{s,i} $ are, respectively, the Einstein A coefficient of excited state $s$ and the atomic transition frequency between the ground and excited state $s$. Furthermore, $I_i$ is the laser intensity on  the $i$th ion. 
When $\Omega_{s,i} \ll \abs{\delta_{s,i}}$, Eq.~(\ref{A1}) is reduced to 
\begin{equation}
    \phi^{\mathrm{opt}}_{s,i} = \frac{\hbar \Omega_{s,i}^2}{4\delta_{s,i}}. 
\label{A3}
\end{equation}

At large laser detunings, the AC Stark shift of the ground atomic level for the $i$th ion follows
\begin{align}
\begin{split}
\phi^{\mathrm{opt}}_{s,i}(\mathbf{r}_i)
={} &  -\frac{3\pi c^2}{2\omega_{s,i}^3} \left( \frac{\Gamma_{s,i}}{\omega_{s,i} -\omega_{l,i}} + \frac{\Gamma_{s,i}}{\omega_{s,i} +\omega_{l,i}} \right) \\
& \times I_i(\mathbf{r}_i - \mathbf{r}^*_i).
\end{split}
\label{A4}
\end{align}
Eq.~(\ref{A3}) and (\ref{A4}) cannot be applied to any real case scenario before accounting for hyper-fine levels of electronic state $s$. For a hyper-fine state labelled by $h$, the transition rate $\Gamma_{h,i} $ is given by 
\begin{equation}
\begin{split}
    \frac{ \Gamma_{h,i} }{ \Gamma_{s,i} } ={}& \frac{ \Omega_{h,i}^2 }{ \Omega_{s,i}^2 } \\
    ={}& (2F_h+1)(2F+1)(2J_h+1) \\
    & \times \left(
    \begin{Bmatrix}
    J_h & J & 1 \\
    F & F_h & I
    \end{Bmatrix}
    \begin{pmatrix}
    F & 1 & F_h \\
    m_F & q & -m_{F,h}
    \end{pmatrix} 
    \right)^2, 
\label{A5}
\end{split}
\end{equation}
where $q = -1, 0, +1$ represent $\sigma^-, \pi, \sigma^+$ polarized light, $J$, $F$, $m_F$, and $J_h$, $F_h$, $m_{F,h}$ are quantum numbers for the corresponding ground and excited hyper-fine state $h$. The AC Stark shift of an electronic state $s$ accordingly follows
\begin{equation}
    \phi^{\mathrm{opt}}_{s,i} = \sum_h \phi^{\mathrm{opt}}_{h,i}. 
\end{equation}

Ions have multiple atomic levels (as for $^{171}$Yb$^+$, see Figure \ref{fig:Yblevels}), which present a more complicated case than a two-level system. In order to find the total AC Stark shift of the ground state of an ion, we can assume that the system is composed of an ensemble of two-level sub-systems consisting of the ground state and each possible excited state. In such a way, the total AC Stark shift of the ground state is the sum of all the AC Stark shifts calculated using different excited states $s$,  
\begin{equation}
    \phi^{\mathrm{opt}}_{i} = \sum_s \phi^{\mathrm{opt}}_{s,i}.
\end{equation}
Here, $\phi^{\mathrm{opt}}_{i}$ can be also regarded as the optical potential energy for the $i$th ion in the ground state. 

\begin{figure*}[t]
\centering
\includegraphics[width=\textwidth]{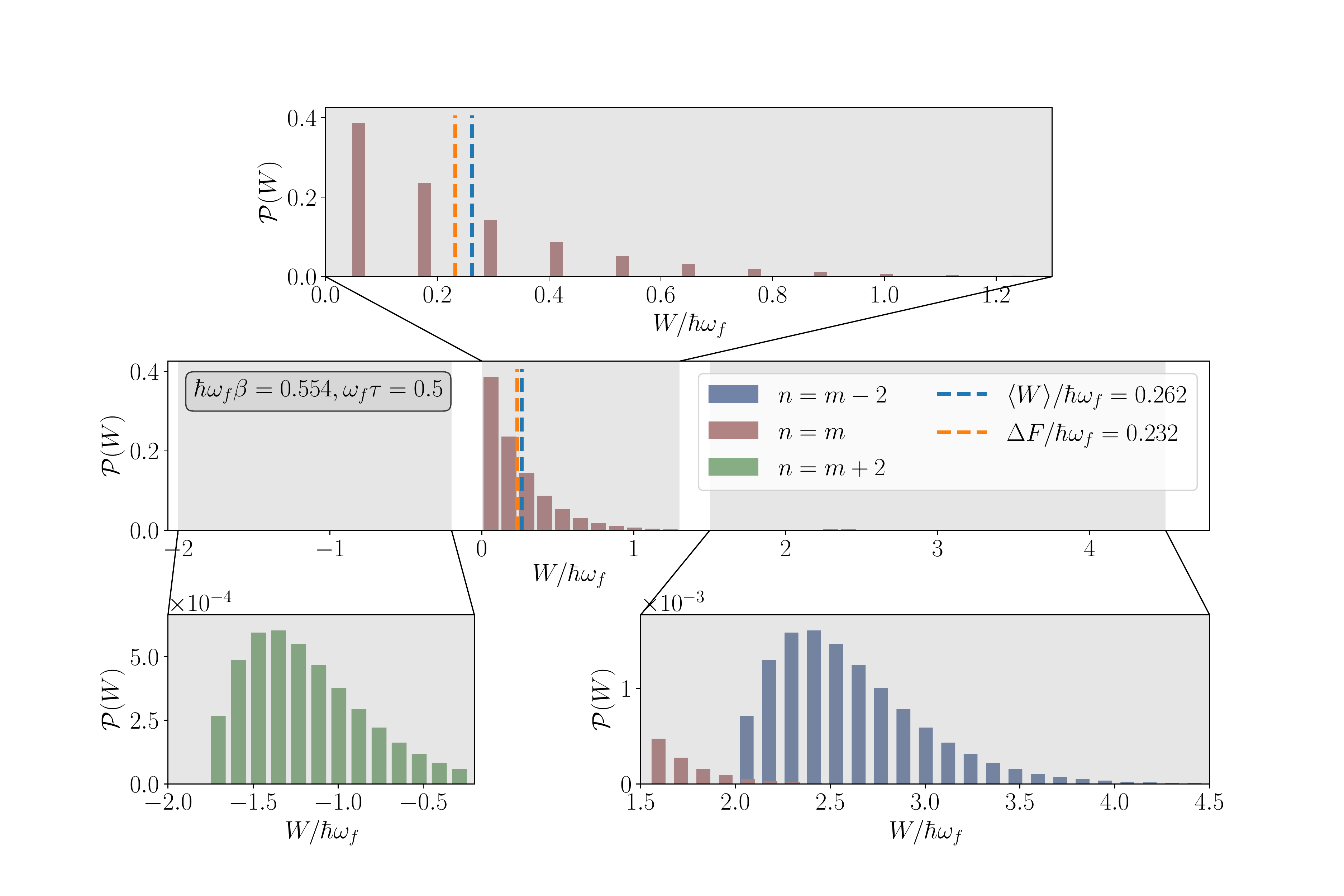}
\caption{
\textbf{Work distribution $\mathcal{P}(W)$ obtained by simulating a $^{133}\mathrm{Ba}^+$ ion with the protocol described in Section~\ref{sec:Thermodynamics}.}
The work distribution $\mathcal{P}(W)$ shown is computed for a temperature of $\hbar\omega_{f}\beta=0.554$ and a switching time of $\omega_{f}\tau = 0.5$.
The various regions of the plot are blown up to account for the differing scale.
The top, lower left and lower right panel each emphasize different kinds of transitions corresponding to $n=m, n=m+2, n=m-2$, respectively.
Only transitions of the kind $n=m \pm 2k \ \text{for} \ k = 0,1 $ are shown as other transitions ($k \ge 2$) have vanishingly small probabilities.}
\label{fig:work_dist}
\end{figure*} 

Furthermore, the scattering rate due to spontaneous emission $r_{\mathrm{sca.}}$ at small laser detunings is given by
\begin{equation}
    r_{\mathrm{sca.},s,i} = \frac{\Omega_{s,i}^2 \Gamma_{s,i}}{ \Gamma_{s,i}^2 + 2\Omega_{s,i}^2 + 4\delta_{s,i}^2}, 
\end{equation}
whereas for large detunings, it follows 
\begin{equation}
\begin{split}
     r_{\mathrm{sca.},s,i} ={}& -\frac{3\pi c^2}{2 \hbar \omega_a^3} \left( \frac{\omega_l}{\omega_{s,i}} \right)^3 I_i(\mathbf{r}_i - \mathbf{r}^*_i) \\
     & \times \left( \frac{\Gamma_{s,i}}{\omega_{s,i} -\omega_l} + \frac{\Gamma_{s,i}}{\omega_{s,i} +\omega_l} \right)^2 . 
\end{split}
\end{equation}
Similar to the calculation of the AC Stark shift, the hyper-fine levels of a state $s$ need to be accounted for using Eq.~(\ref{A5}) and we have
\begin{equation}
    r_{\mathrm{sca.},s,i} = \sum_h r_{\mathrm{sca.},h,i}, 
\end{equation}
and
\begin{equation}
    r_{\mathrm{sca.},i} = \sum_s r_{\mathrm{sca.},s,i}. 
\end{equation}

Finally, all the low-lying states, as compared to the visible and near-infrared transitions of $^{171}$Yb$^+$, which are electric-dipole allowed to couple to the ground state $S_{1/2}$ are shown in Figure \ref{fig:Yblevels}. Note, $D$ and $F$ manifolds of $^{171}$Yb$^+$ are not included in the figure. It should be also noted that in the calculation of optical trapping parameters (see Table \ref{tab:optparams}) all the hyper-fine levels associated with different states shown in Figure \ref{fig:Yblevels} are taken into consideration.

\section{Work distribution}
\label{apdx:workdist}

In Section~\ref{sec:Thermodynamics}, we discussed the use of Jarzynski's equality to extract the free energy difference between $^{133}\mathrm{Ba}^+$ and $^{171}\mathrm{Yb}^+$ ion from the work distribution $\mathcal{P}(W)$ (Eq.~\ref{pW_defn_eqn}).
Fig.~\ref{fig:work_dist} explicitly shows $\mathcal{P}(W)$ as a multi-modal distribution with a peak for each value of $k$, where $n=m \pm 2k$.

% ================
% End Appendix
% ================

\end{document}